# Block size estimation for data partitioning in HPC applications using machine learning techniques

Riccardo Cantini[a], Fabrizio Marozzo[a,b], Alessio Orsino[a], Domenico Talia[a,b], Paolo Trunfio[a,b], Rosa M. Badia[c], Jorge Ejarque[c], Fernando Vazquez-Novoa[c]

[a] *University of Calabria*
[b] *DtoK Lab Srl*
[c] *Barcelona Supercomputing Center (BSC)*

**Abstract**

The extensive use of HPC infrastructures and frameworks for running data-intensive applications has led to a growing interest in data partitioning techniques and strategies. In fact, application performance can be heavily affected by how data are partitioned, which in turn depends on the selected size for data blocks, i.e. the block size. Therefore, finding an effective partitioning, i.e. a suitable block size, is a key strategy to speed-up parallel data-intensive applications and increase scalability. This paper describes a methodology, namely BLEST-ML (BLock size ESTimation through Machine Learning), for block size estimation that relies on supervised machine learning techniques. The proposed methodology was evaluated by designing an implementation tailored to *dislib,* a distributed computing library highly focused on machine learning algorithms built on top of the PyCOMPSs framework. We assessed the effectiveness of the provided implementation through an extensive experimental evaluation considering different algorithms from dislib, datasets, and infrastructures, including the MareNostrum 4 supercomputer. The results we obtained show the ability of BLEST-ML to efficiently determine a suitable way to split a given dataset, thus providing a proof of its applicability to enable the efficient execution of data-parallel applications in high performance environments.

## 1. Introduction

Data partitioning refers to splitting a dataset into small and fixed-size units, called blocks or chunks, to enable efficient data-parallel processing and storing in distributed-memory-based systems. Several issues related to data partitioning must be addressed to reduce execution times and ensure the good scalability of applications. For example, when a dataset is mapped on a set of nodes of a parallel/distributed computing system, two very critical problems are highlighted: (i) the choice of the destination node for a given block (i.e., the node where that block will be stored); and (*ii*) the selection of an appropriate size of the blocks the dataset is divided into, i.e. the block size. The first problem has been addressed in several studies [1, 2, 3, 4], in which scheduling algorithms have been proposed to minimize data movement at run-time. The second problem addressed in this work has been tackled in the literature by different distributed computing frameworks, through fixed-size partitioning strategies and heuristics [5, 6]. Besides the use of built-in partitioning techniques provided by these frameworks, general-purpose autotuners [7] can be exploited to adjust several hyperparameters, including the block size.

The block size can heavily affect the trade-off between single-node efficiency and parallelism in data-intensive applications. Specifically, a larger size reduces parallelism (fewer blocks) but makes tasks larger. Although this can lead to an overhead reduction, it must be ensured that the block size does not exceed the memory available on the individual nodes, so as to avoid memory saturation. On the other hand, a smaller size leads to finer exploitation of parallelism, while introducing a larger overhead due to communication, synchronization, and task management, which can negatively impact performance.

In this work, we propose BLEST-ML (BLock size ESTimation through Machine Learning), a novel methodology for data block size estimation that relies on supervised machine learning techniques. Specifically, it leverages a cascade of tree-based classifiers to determine a proper value for the block size, given an execution to be performed. The model is trained on a log of past executions, represented by a set of descriptive features, including algorithm, dataset, and execution environment characteristics.

The effectiveness of BLEST-ML was assessed by designing an implementation tailored to *dislib*, a distributed computing library focused on machine learning algorithms. We carried out an extensive experimental evaluation on different execution environments, including the MareNostrum 4 supercomputer (MN4) [8] located at the Barcelona Supercomputing Center (BSC), and several real-world datasets, including High Energy Physics (HEP) and bio-informatics. The results show the ability of the proposed methodology to efficiently predict a suitable value for the block size in dislib applications running on a large-scale High Performance Computing (HPC) system such as MN4. Our findings provide a proof of the applicability of BLEST-ML to support programmers in choosing proper data partitioning, thus enabling the efficient execution of data-parallel applications in HPC environments.

The main contribution of the proposed work lies in the application of supervised machine learning techniques for block size estimation, facilitating efficient data partitioning in HPC applications. Specifically, we employ a cascade of decision models trained on a historical log of past executions to learn the patterns that connect a given configuration to the most suitable block size. Adopting this approach, BLEST-ML can determine a good estimate in a quick and efficient way, demanding minimal domain knowledge, and avoiding the necessity for resource-intensive exploration of vast search spaces and dynamic profiling at runtime. Furthermore, it can handle non-monotonic relationships between performance and configuration, which makes it applicable in a wide range of use cases related to the execution of data-parallel applications in distributed and HPC environments. All these factors collectively make our approach more efficient and effective than the existing trial-and-error heuristics, and a valuable alternative to autotuning frameworks.

The remainder of the paper is organized as follows. Section 2 discusses the main methods for data partitioning present in the literature. Section 3 describes the proposed methodology. Section 4 introduces the dislib library of PyCOMPSs, i.e. the framework chosen as the testbed. Section 5 presents the experimental evaluation and results. Finally, Section 6 concludes the paper.

### *1.1. Problem statement*

The problem of partitioning a dataset in a distributed environment requires defining how the dataset should be divided into blocks. Besides some techniques that use fixed-size blocks, many others leverage the concept of block size that defines the number of rows and columns of each block the dataset is divided into. The choice of this parameter is challenging but key to speed-up parallel data-intensive applications and increase scalability, as it determines the trade-off between parallelism, scheduling overhead, and memory usage.

Typically, block size estimation is not an easy task for programmers. In fact, they usually proceed by following a trial-and-error approach, only supported by simple heuristics and domain knowledge (i.e., the awareness of the behavior of the algorithm in a given distributed environment). As a result, this tuning process is often time-consuming and resource-intensive, especially when large datasets and complex hardware infrastructures are used.

The block size estimation problem is formulated as follows. Let $d$ be a dataset composed of $n$ rows and $m$ columns, and $a$ the algorithm to be run on $d$ within the execution environment $e$. The goal is to determine the best size of data blocks, that is a bidimensional variable $(r^*, c^*)$, in which $r^*$ and $c^*$ represent the optimal number of rows and columns of each block. A slightly different approach, used in this work, formulates the problem as the prediction of the number of rows and columns partitions in which to divide the dataset, i.e. the target variable $(p_r^*, p_c^*)$. Starting from this, the block size is then obtained as $(r^*, c^*) = (n/p_r^*, m/p_c^*)$. For the sake of clarity, Table 1 reports the meaning of the main symbols used throughout the paper.

| Symbol | Meaning |
|---|---|
| $d$ | Dataset composed by $n$ rows and $m$ columns. |
| $e$ | A representation of the target execution environment. |
| $a$ | The machine learning algorithm to be executed on $d$ within $e$. |
| $p_r^*$ | The optimal number of partitions along rows. |
| $p_c^*$ | The optimal number of partitions along columns. |
| $r^*$ | The optimal number of rows in a block. |
| $c^*$ | The optimal number of columns in a block. |
| $(r^*, c^*)$ | The optimal block size expressed as $(n/p_r^*, m/p_c^*)$. |

Table 1: Meaning of the main symbols used throughout the paper.

## 2. Related work

In high performance computing, data partitioning is a key strategy to speed-up parallel data-intensive applications and increase scalability. In this section, we describe the main methods proposed in the literature and used in HPC infrastructures.

**Horizontal, vertical, and hybrid partitioning.** The data partitioning problem is generally addressed by using three main alternative techniques: horizontal, vertical, or hybrid partitioning [9].

- *Horizontal partitioning.* This technique, also called sharding, divides the rows of the dataset into disjoint subsets so that each subset has the same number of columns as the whole dataset. This approach is used in many state-of-the-art frameworks for big data processing, such as Hadoop and Spark [6, 5].
- *Vertical partitioning.* In this approach, the columns of the dataset are divided into subsets, usually based on the columns on which data querying is more frequently performed, or a heuristic approach [10, 11].
- *Hybrid partitioning.* It combines the horizontal and vertical approaches, by aggregating data according to how it is used by the target application and/or system [12, 13]. It is also called functional partitioning.

Among all described strategies, horizontal partitioning is the one that is mostly used in big data applications and HPC systems, while the other two strategies are less used and therefore explored in the literature. Specifically, the horizontal partitioning can be further categorized into *hash-*, *range-*, and *random-* based. The hash-based approach divides records into subsets by hashing the record key and then mapping the hash value of the key to a partition. A common method to do this mapping is using a round-robin algorithm, which guarantees a balanced partitioning among nodes and partitions of equivalent size. In hash-based partitioning records having the same key value must have the same hash value, and consequently, they will be mapped to the same partition. The range-based partitioning approach, instead, partitions a dataset according to a given range and distributes records having the keys within the same range on the same node. In distributed environments, how to set this range is often challenging, especially when dealing with large-scale data analysis. Finally, the random-based partitioning approach divides the records randomly into subsets using a random number generator to determine where to distribute each record, producing approximately subsets of equal size. As an example, Salloum et al. proposed the Random Sample Partition (RSP) data model to support distributed big data analysis [14, 15]. This model represents a big dataset as a set of non-overlapping blocks, where each block is a random sample of the whole dataset. In addition, Wei et al. [16] proposed a two-stage algorithm to generate RSP data blocks from the Hadoop Distributed File System (HDFS). A drawback of random-based approaches is that they do not consider the correlation between records, which instead can be leveraged to avoid unnecessary computations. As an example, in [17] the authors designed a context-based multi-dimensional partitioning technique that relies on data correlation to determine a suitable split.

**Static and dynamic partitioning.** Main approaches for data partitioning can be further categorized into *static* or *dynamic* [18]. Static approaches use a fixed size when a block is selected, often defined in MB, and the partitioning is computed before starting the execution. As an example, data in HDFS is divided into fixed-size blocks, obtained from horizontal partitioning. Particularly, a typical block size is 128 MB, which means that if we have a 1 GB file, it will be partitioned into 8 blocks, each one of 128 MB. Similarly, in Spark, the HDFS blocks need to be loaded into an in-memory data structure, called Resilient Distributed Dataset (RDD) [19], which can be then partitioned using either the aforementioned strategies, such as hash and range, or a custom partitioning. Specifically, Spark runs a single task for every partition of an RDD, up to $2-3$x times the number of cores in the cluster. Hence, a heuristic can be derived that determines the number of partitions as a small multiple of the total number of available cores. On the contrary, in the dynamic approaches, the dimension and shape of the blocks in which to partition the dataset are not chosen a priori, but at runtime. As an example, several dynamic strategies can be found in [18], specially designed for graph data partitioning. Both approaches present some issues. Static approaches define a priori how data should be partitioned, without taking into account any dataset characteristic or algorithm feature. On the other hand, dynamic approaches support adaptation to the actual workload at runtime, but they can introduce significant overhead in dynamically adjusting data partitioning. To overcome these issues, in this work we propose a hybrid and static data partitioning approach, able to determine how to adequately partition the dataset before the execution, thus avoiding the overhead introduced by dynamic approaches, while also taking into account dataset and algorithm characteristics and infrastructure features. Specifically, we determine the best size of data blocks by using a supervised machine learning technique, focusing on the estimation of two quantities, i.e. the number of partitions in which to divide dataset rows and columns. This allows finding a good partitioning to make the most of the computational resources of the execution environment, thus improving application performance and scalability.

**Autotuners.** Besides the discussed techniques specially designed to address the data partitioning problem, general-purpose *autotuners* can be used to tune different application parameters, including the block size, in order to improve application performance and throughput. As an example, OpenTuner [7] is a framework for building domain-specific multi-objective program autotuners, which leverages an ensemble of search techniques that can be run simultaneously. Candidate configurations (i.e., selected points in a user-defined multi-dimensional search space) are evaluated through a meta-search strategy, which allocates tests to search techniques, relying on the resolution of the multi-armed bandit problem. Among the other tools, SmartConf [20] relies on control theory to optimize system performance and stability, by performing automated and dy-namic configuration adjustments, based on performance constraints. Although the aforementioned tuners represent general-purpose solutions, which can be effectively used for parameter configuration, they have some limitations. As an example, SmartConf suffers from the lack of capacity to deal with the non-monotonic relationship between performance and configuration. Specifically, considering the block size estimation problem - on which the present work is focused - there exists a non-monotonic relationship between the number of blocks and application performance since too few blocks would hinder application parallelism, while too many blocks would introduce a non-negligible overhead which results in significant performance degradation. In such a case, as stated by the authors in [20], a machine-learning solution - like the one proposed in this work - would be a better fit. Such a solution, in addition, can also be faster than autotuners, by quickly providing a suggestion that solely relies on the learned patterns linking a given configuration with the most suitable block size, without the need for resource-intensive exploration of vast search spaces and dynamic profiling at runtime. Nonetheless, it is worth noticing that solutions like OpenTuner [7] cope with this issue by leveraging intelligent search mechanisms aimed at evaluating a relatively small set of candidate configurations, thus effectively handling huge search spaces whose full exhaustive search would be unfeasible. Furthermore, machine learning-based approaches may require a large amount of historical training data in order to generalize well to new or unseen inputs, and are generally tailored to a specific optimization problem.

## 3. A machine learning approach for block size estimation

BLEST-ML leverages a classification-based approach to address the problem of data partitioning, by determining the best block size $S = (n/p_r, m/p_c)$ that minimizes the execution time of algorithm $a$ on a dataset $d$ composed of $n$ rows and $m$ columns. Specifically, the target classes to be predicted by the machine learning model are $p_r$ and $p_c$, two discrete variables ranging from 1 to a maximum number of partitions, usually defined as a small multiple of the total number of cores available. We selected this kind of approach because it results to be more stable compared to a regression-based one, in which the block size is directly predicted by identifying the number of elements of each block. In fact, the main problem of the regression-based approach is that its output is generally unconstrained, which may lead to a set of blocks with a non-uniform size. The classification-based approach, instead, is less affected by this problem, as it selects the number of partitions against a finite number of possible values. However, the ability to generalize heavily depends on the representativeness of the training data, which implies that the produced estimates are reliable for HPC systems having similar infrastructure features and datasets whose size is of the same order of magnitude as those seen during training. In the following, we provide a detailed description of the three main steps that make up BLEST-ML (see Figure 1).

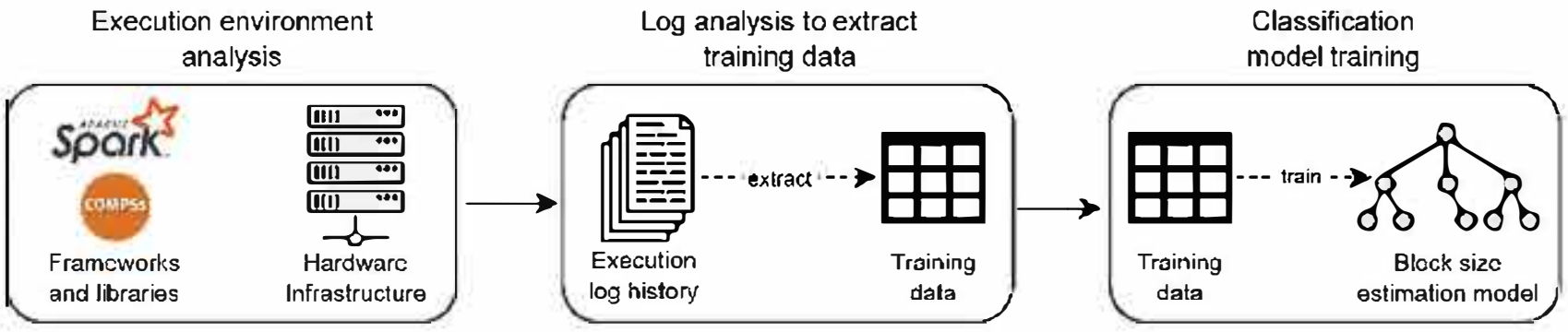

Figure 1: Execution flow of BLEST-ML (BLock size ESTimation through Machine Learning).

### 3.1. Execution environment analysis

Given a distributed environment in which data analysis applications can be run, BLEST-ML aims to enable their efficient execution by identifying a proper size for data blocks. This can help programmers to make the most of all the computing and storage resources that are available in the environment, as they can efficiently obtain a suitable estimate for the block size, without the need for heavy tuning processes or domain knowledge. As a preliminary step, the execution environment must be carefully analyzed. Generally, it is characterized by a set of software features, such as the available frameworks and libraries, and infrastructure features, such as the number of nodes, cores per node, available memory, and disk space.

### 3.2. Log analysis to extract training data

BLEST-ML leverages a log of past executions to extract the patterns that link a specific execution to the best block size, by training a supervised machine learning model. However, in order to learn effective patterns, raw logs must be adequately processed to extract an appropriate set of training data. The log $\mathcal{L}$ consists of a collection of executions, performed by both standard users and domain experts, in which a single execution is described by the characteristics of the dataset ($d$), the algorithm ($a$), the execution environment ($e$), the applied partitioning along rows ($p_r$) and columns ($p_c$), the overall execution time ($t$), and other measurements such as main memory/disk usage. Formally, an execution in $\mathcal{L}$ is represented by the tuple $\langle d, a, e, p_r, p_c, t \rangle$. Based on the information available in $\mathcal{L}$ and the application domain, which includes the execution framework and the target infrastructure (e.g., a single node or an HPC cluster), $d, a$ and $e$ can be represented in different ways to consider domain-specific aspects. As an example: *i*) for the dataset $d$, the number of rows and columns can be considered, together with the dataset size that is useful when a fixed- size partitioning is used; *ii*) for the algorithm $a$, it could be relevant to discern between the type of task (e.g., classification or clustering) and usage mode (i.e., training and inference); *iii*) for the execution environment $e$, the number of cores and the available amount of RAM can be used, while other features like the number of nodes and the RAM per node only apply to distributed infrastructures.

In order to generate the training dataset $\mathcal{D}$, all executions in $\mathcal{L}$ are grouped by the triple $\langle d, a, e \rangle$. In this way, we can observe how, given a fixed configuration, execution time is affected by different block sizes. Afterward, for each group, the best partitioning ($p_r^*$, $p_c^*$) that led to the minimum execution time is found and the tuple $\langle d, a, e, p_r^*, p_c^* \rangle$ is added to $\mathcal{D}$. At the end of the process, the dataset $\mathcal{D}$ will contain the best partitioning found for each triple $\langle d, a, e \rangle$, specifically:

- Features related to algorithm $a$, dataset $d$ (dimension in MB, number of rows, etc.), and execution environment $e$ (number of cores, number of nodes, etc.).
- The optimal partitioning ($p_r^*, p_c^*$), i.e. the target variable.

Table 2 shows an example of an excerpt of $\mathcal{D}$ obtained from the training data extraction step.

| *Algorithm* | *Dataset rows* | *Dataset columns* | *Dataset size (GB)* | *Infrastructure features* | | | *Best partitioning* | |
|---|---|---|---|---|---|---|---|---|
| | | | | *# nodes* | *# cores* | *RAM* | $p_r^*$ | $p_c^*$ |
| K-means | 500,000 | 1000 | 2.39 | 4 | 64 | 256 | 32 | 4 |
| Random Forest | 1000 | 500,000 | 2.92 | 4 | 64 | 256 | 32 | 8 |
| SVM | 10,000 | 10,000 | 1.1 | 4 | 64 | 256 | 16 | 16 |

Table 2: Excerpt of the training set extracted by the log of executions.

An information-rich log, from which to extract a fairly representative dataset, is generally available, as current distributed processing frameworks, such as PyCOMPSs and Apache Spark, provide accurate instrumentation tools for collecting a wide range of information, which is usually stored with the aim of facilitating application performance monitoring. However, depending on the particular use case, it may be required to integrate D with supplementary executions to ensure high-quality estimates. To face this issue, training data can be generated and/or enriched by arranging a set of executions, with the aim of finding the block size that optimizes execution time for the considered configurations. This process is characterized by several degrees of freedom, including the executed algorithm ($a$), input data characteristics ($d$), and infrastructure features of the execution environment ($e$), which leads to the need for an efficient search strategy. For this purpose, a grid search technique can be leveraged, in which several triples $\langle d, a, e \rangle$ are generated and annotated with the best block size found during the search. Specifically, for each triple, the following operations are performed.

- Given $n_{cores}$ the number of available cores, a $k \times k$ grid $\mathcal{G}$ is built, with $k = log_s(n_{cores})$, where $s$ is a search step such that $log_s(n_{cores})$ is an integer number. The step $s$ (set to 2 by default) can be used to control the trade-off between the cost of the grid search and the representativeness of the generated training samples.
- Each element $g_{i,j}$ in the grid $G$, with $i$ and $j$ ranging from 1 to $k$, is determined as the time of executing algorithm $a$ on the dataset $d$ within the environment $e$, by splitting $d$ using the ($p_r = s^i$, $p_c = s^j$) partitioning.

This means that the rows and columns of $d$ will be divided into $s^i$ and $s^j$ blocks, respectively. The execution time in the case of failures (e.g., out-of-memory errors) is set to $\infty$.

- By exploring the grid, the best partitioning $(p_r^*, p_c^*)$ for the triple $\langle d, a, e\rangle$ is found, which leads to the minimum execution time. Formally, it is computed as the pair $(p_r^*, p_c^*) = (s^{i^*}, s^{j^*})$, where $(i^*, j^*) = \arg\min_{i,j} \mathcal{G}$. Finally the triple $\langle d, a, e\rangle$ is labeled with $(p_r^*, p_c^*)$ and added to the training dataset $\mathcal{D}$.

It is important to notice that these supplementary executions occur offline (i.e., at any time before an actual prediction is required). For instance, these additional executions can be scheduled as low-priority jobs. Consequently, the time required to complete these executions will not affect the response to a user query (i.e., a suitable block size for the submitted application). In this way, the time-to-solution can be drastically reduced compared to manually trying several possible configurations when a request is submitted by the user.

This approach for training data generation via execution monitoring was successfully leveraged in [21], whose purpose is to improve the in-memory execution of data-intensive workflows on parallel machines [22].

### *3.3. Classification model training*

Given the dataset $\mathcal{D}$ obtained in the previous step, a classification model is trained to learn the patterns that relate the execution features/parameters and the best partitioning $(p_r^*, p_c^*)$. Thus, the output of this step is a classification model capable of estimating the optimal number of partitions in which to split the rows and the columns of a given dataset, based on its characteristics, the algorithm to be run, and the underlying execution environment. Since the target variable to be predicted, that is the best partitioning $(p_r^*, p_c^*)$, is two-dimensional, a multi-output classification model is needed. Among the main approaches for multi-output classification, a popular one consists in fitting a separate classifier for each dimension of the target variable (i.e., two separate classifiers in our case). The main drawback of such an approach is that it ignores the relationships between the predicted outputs of the single classifiers. For this reason, a stacking approach is leveraged that allows considering such relationships by using a cascade of two different decision tree classifiers. The two classifiers, namely $DT_r$ and $DT_c$, are used to predict the best number of rows and column blocks, respectively. This approach is more suited for the case of block size, in which the number of rows and the number of columns in a block are very likely to be dependent on each other. Therefore, in the chained model, the predictions of $DT_c$ are conditioned on the output of $DT_r$, as shown in Figure 2. In addition, we followed this order in chaining the two decision tree models since partitioning along the rows is generally more relevant. The training step is performed as described below.

1. We train the first decision tree $DT_r$ with the training instances of $\mathcal{D}$ to learn the number of blocks $p_r^*$ in which to partition dataset rows.

2. The second decision tree $DT_c$ is trained with the training instances concatenated with the output of $DT_r$, with respect to the second target variable $p_c^*$, to learn the number of blocks in which to partition dataset columns.

Afterward, the corresponding block size is determined as $(r^*, c^*) = (n/p_r^*, m/p_c^*)$, where $n$ and $m$ are the rows and columns of the considered dataset.

For the sake of clarity, we report an end-to-end example of computing the block size for a given input instance. Let $n = 51,200$ and $m = 256$ be the number of dataset rows and columns, respectively. Suppose we have to predict the block size related to the execution of an SVM algorithm, and that the result of the prediction is $(p_r^*, p_c^*) = (4, 16)$. Then the optimal block size can be computed as follows:

$$(r^*, c^*) = (n/p_r^*, m/p_c^*) = (12800, 16)$$

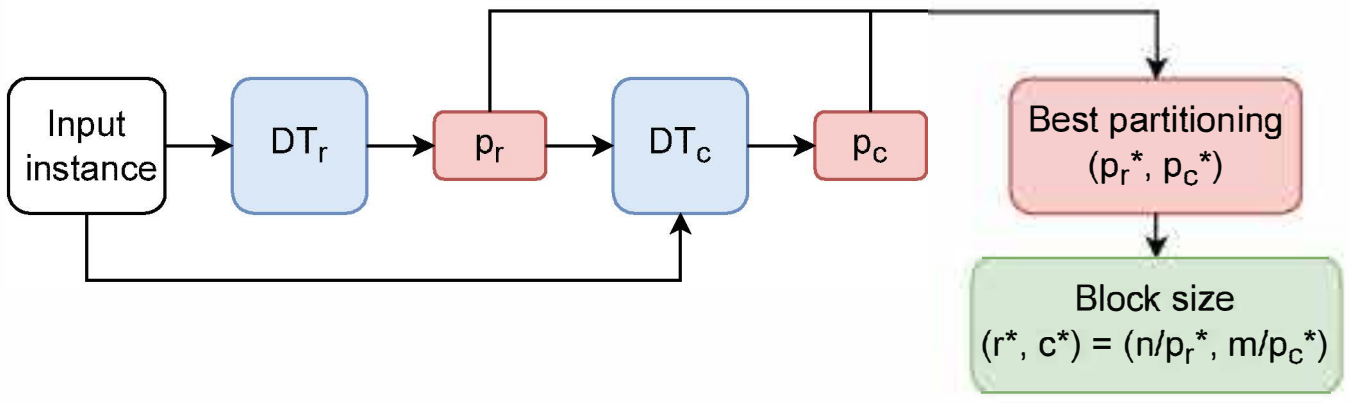

Figure 2: Chained multi-output classification model.

## 4. Block size estimation in dislib applications

The proposed methodology can be applied to a wide range of frameworks for distributed data processing [23]. In fact, the majority of these systems, such as Hadoop [6], Spark [5], DMCF [24] and PyCOMPSs [25], leverage a data-parallel approach that involves a data partitioning step for distributing the dataset across a set of working nodes. Consequently, BLEST-ML can bring huge benefits, by suggesting an adequate partitioning that allows to effectively run distributed applications, reducing overhead while ensuring a good level of parallelism and throughput. Among the main frameworks and libraries of interest for distributed data processing, we selected as a testbed PyCOMPSs, focusing on dislib [26], a distributed computing library built on top of it that provides distributed machine learning algorithms. The early implementation, based on PyCOMPSs and dislib, is publicly available on GitHub[1].

PyCOMPSs [25, 27] is a task-based programming model that enables the parallel execution of sequential Python code in distributed computing platforms.

[1]https://github.com/eflows4hpc/dislib-block-size-estimation

By means of Python decorators, the developer identifies the function/methods to be considered tasks. PyCOMPSs also offers a small API for synchronization. It is based on a runtime able to identify the data dependencies that exist among tasks building a data dependency graph. The task graph exposes the possible task concurrency that is exploited by the runtime, which manages the execution of the tasks in distributed infrastructures, scheduling them, and performing all the necessary data transfers. A task in PyCOMPSs can run using multiple cores if internally parallelized with threads or other alternative programming models such as OpenMP.

The Distributed Computing Library (dislib) [26], implemented on top of PyCOMPSs, provides various distributed algorithms for several machine learning tasks, including classification, clustering, and dimensionality reduction. Dislib is inspired by scikit-learn and NumPy, and it comes with two primary programming interfaces: an API to manage data in a distributed way and an estimator-based interface to work with different machine learning models. Its main data structure is the distributed array (*ds-array*) which enables it to distribute the datasets in multiple nodes of a distributed infrastructure. A ds-array is a matrix divided into blocks, which can be a NumPy array or a SciPy CSR (Compressed Sparse Row) matrix, depending on the kind of data used to create the ds-array. Dislib provides an API similar to NumPy to work with ds-arrays, but ds-arrays are stored remotely, allowing to store much more data than regular NumPy arrays. All operations on ds-arrays are internally parallelized with PyCOMPSs. The typical workflow in dislib consists of the following steps: *i*) reading input data into a ds-array; *ii*) creating an estimator object; *iii*) fitting the estimator with the input data; *iv*) getting information from the model's estimator or applying the model to new data. At each step, the level of parallelism is driven by the number of blocks of the ds-arrays that are operated, which in turn is controlled by the ds-array's block size, which defines the number of rows and columns of each block.

Choosing the right size of a block-array can be a quite challenging task: small blocks allow for higher parallelism as the computation is divided into more tasks. However, handling a large number of blocks can generate overhead that can negatively impact performance. Thus, the optimal block size will allow the full utilization of the available resources without adding too much overhead. In addition to this, block size also affects the amount of data that tasks load into memory. This means that block size should never be bigger than the amount of available memory per processor. Summing up, the choice of the optimal block size is often difficult but essential for exploiting the full potential of dislib, hence the possibility of effectively applying the proposed methodology.

## 5. Experimental evaluation

This section presents the extensive experimental evaluation we carried out to assess the effectiveness of BLEST-ML, analyzing how the partitioning suggested by BLEST-ML can improve the execution of dislib applications in different scenarios. Specifically, we evaluated our methodology in a single-node and a multi-node execution environment, i.e. a cluster node and the MareNostrum 4 supercomputer.

For what concerns the evaluation metrics, we used *makespan ratio* to measure the improvement in speed of execution brought by the predicted block size, with respect to other possible partitions. Specifically, given an algorithm $a$ to be run in a distributed environment $e$, let $t^*$ and $t_{other}$ be the execution times achieved by using the predicted block size and a different one, respectively. We compute the *makespan ratio* as follows:

$$makespan\ ratio = \frac{t_{other}}{t^*}$$

In addition, we measured the percentage *makespan reduction*, i.e. the percentage amount of execution time saved by running a given algorithm with the predicted block size, with respect to a different one. Formally:

$$makespan\ reduction = \frac{t_{other} - t^*}{t_{other}}$$

### 5.1. Single-node experiments

The used log contains information about almost 5000 executions performed on datasets of varying sizes using a wide range of machine learning algorithms provided by dislib for classification and clustering, including Support Vector Machine (SVM), Random Forest (RF), Gaussian Mixture Model (GMM), and K-means. In the following sections, we describe the results achieved in the single-node scenario, by evaluating the benefits brought by the estimated block size with the use of both real-world and synthetic test datasets.

#### 5.1.1. Real-world datasets

The effectiveness of BLEST-ML in suggesting a suitable block size value was evaluated on two real-world datasets used for clustering and classification:

- *HEPMASS* [28]: it is a high-energy physics dataset containing signatures of exotic particles, learned from Monte Carlo simulations of the collisions that produce them. The dataset contains 7 million training samples with 27 features that can be separated into two clusters, i.e. particle-producing collisions and background sources.
- *MNIST* [29]: it is a multi-class dataset used for image classification and pattern recognition, containing gray-scale images of handwritten digits, from 0 to 9, labeled with the represented number. In particular, the dataset contains 60 thousand training images in a $28 \times 28$ format, which can be represented by vectors of 784 features.

Since both test datasets are characterized by a big number of rows against a relatively small number of columns, BLEST-ML suggested in both cases a block size that partitions both datasets only horizontally, that is just one block for the columns containing all features. For this reason, the number of blocks generated by creating the distributed arrays is equal to the number of partitions along rows. The results achieved by running K-means and Random Forest on HEPMASS and MNIST datasets are summarized in Table 3.

| *Algorithm* | *Dataset name* | *Dataset rows* | *Dataset columns* | *Metric* | *Best time* | *Average time* | *Worst time* |
|---|---|---|---|---|---|---|---|
| K-means | HEPMASS | $7 \cdot 10^6$ | 27 | Makespan ratio<br>Makespan red. | $0.96 \pm 0.03$<br>$-3.80\% \pm 0.09$ | $1.48 \pm 0.04$<br>$32.6\% \pm 0.05$ | $2.53 \pm 0.07$<br>$60.5\% \pm 0.06$ |
| Random Forest | MNIST | $6 \cdot 10^4$ | 784 | Makespan ratio<br>Makespan red. | $1.00 \pm 0.01$<br>$0\% \pm 0.01$ | $1.27 \pm 0.03$<br>$21.32\% \pm 0.04$ | $1.65 \pm 0.03$<br>$39.51\% \pm 0.06$ |

Table 3: Makespan ratio and percentage makespan reduction measured by running K-means and Random Forest algorithms on the HEPMASS and MNIST datasets.

Specifically, the time $t^*$ achieved by running the two algorithms using the predicted block size was compared against the *best*, *worst*, and *average* times achieved by using all other possible partitionings, calculated using progressive powers of 2 from 2 to 256, i.e. 4x times the total number of cores available. The best time provides an upper bound to $t^*$, while the worst time is used as a lower bound for performance. Moreover, Figure 3 shows the measured execution time by using different gradations of red, where a greater intensity corresponds to a higher duration. In the proposed plots, the time obtained by using the predicted partitioning is marked by a cyan circle, while the best one is marked by a green star.

By observing Figure 3(a), it can be noticed that the optimal number of blocks that led to the best execution time for the K-means algorithm was 16, while BLEST-ML suggested partitioning rows in 32 blocks. However, the time measured by using the predicted block size, i.e. $t^*$, is the second best time, and the difference with the best one is negligible ($\approx$ 1 second). By comparing $t^*$ with the average execution time, the predicted block size led to a 1.48 makespan ratio, with a percentage reduction of makespan equal to 32.6%. The worst execution time was observed in the case in which 256 blocks were used. This is caused by the excessively small size of the blocks, which leads to the generation of a too large number of blocks and tasks. In fact, such a degree of parallelism produces too much overhead that results in a degradation of application performance. A similar execution time was measured when just one block was used, i.e. no partitioning is performed. This is the opposite case, in which parallelism is not exploited at all. By comparing $t^*$ with the worst execution time, we measured a makespan ratio of 2.53 and a makespan reduction of 60.5%, which confirms how the block size suggested by BLEST-ML allows determining a proper partitioning, which leads to a quite good trade-off between the degree of parallelism and the introduced overhead. The quality of the partitioning suggested by BLEST-ML is further confirmed by the execution of Random Forest on the MNIST dataset (Figure 3(b)). In this case, BLEST-ML predicts exactly the best possible partitioning, i.e. 16 blocks along rows. Also in this case, the worst values were measured at the extremes, where the level of parallelism is either zero (1 block) or too high (256 blocks). Furthermore, we measured a makespan ratio of 1.27 and 1.65 and a makespan improvement of 21.32% and 39.51%, compared to the average and worst execution times, respectively.

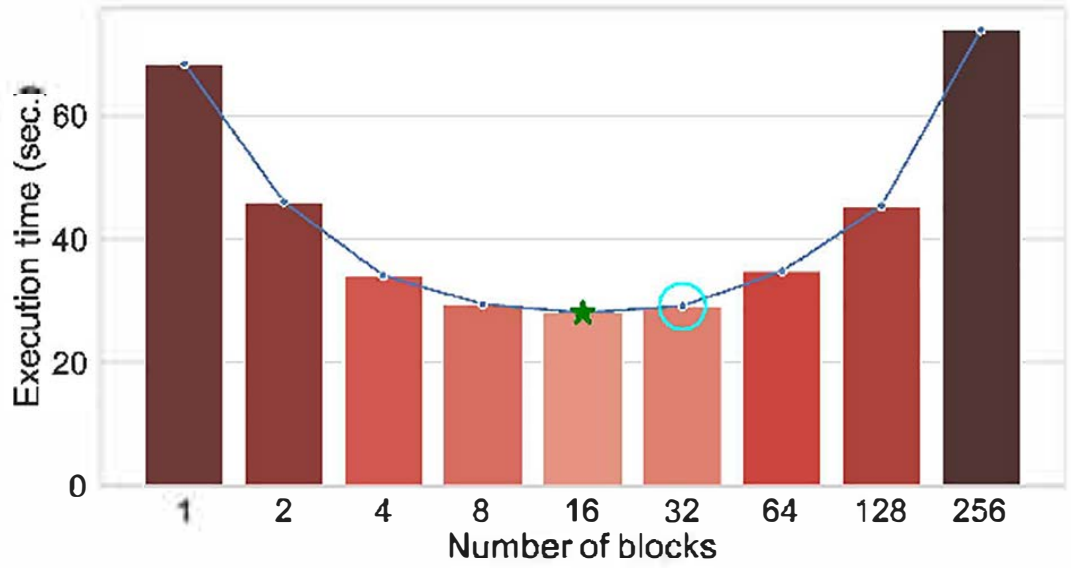

(a) K-means, HEPMASS dataset.

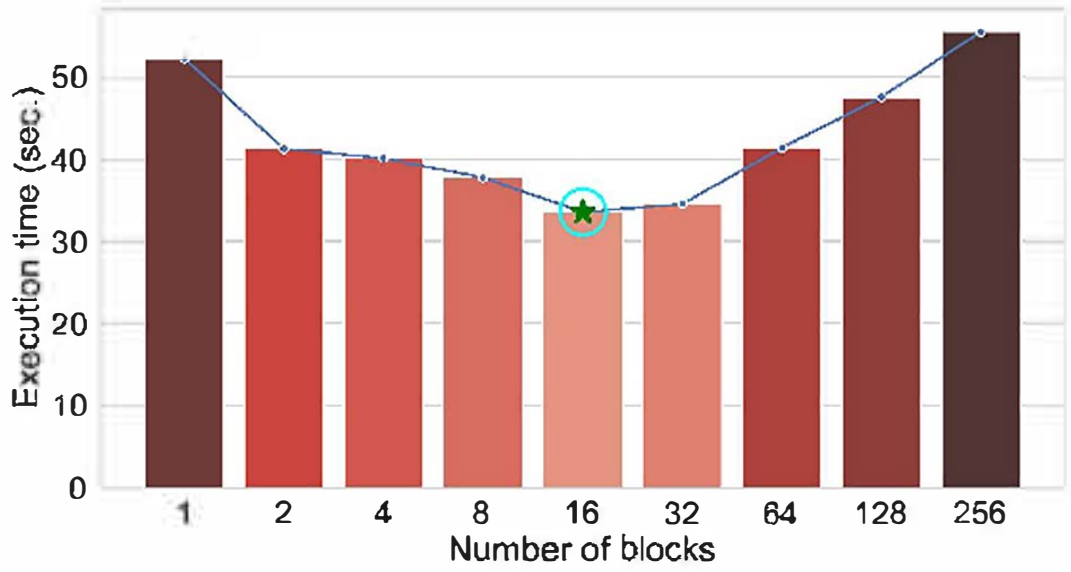

(b) Random Forest, MNIST dataset.

Figure 3: Execution times (in seconds) achieved by K-means and Random Forest executed on two real-world datasets. The time obtained by using the predicted number of row blocks is marked by the cyan circle, while the best one is marked by the green star.

#### *5.1.2. Synthetic datasets*

To further explore the effectiveness of BLEST-ML, its estimates were evaluated against a set of synthetic test datasets, which is useful to observe how the algorithms behave in some specific cases. For this purpose, we generated a series of multiclass test datasets, by allocating one or more normally-distributed clusters of points to each class. Particularly, we used both isotropic and anisotropic Gaussian blobs. In addition, the obtained samples were augmented with random noise and redundant features, generated as a linear combination of the original ones.

Starting from a set of synthetic test datasets of varying shapes, generated following the aforementioned process, we measured the performance improvement achievable with the use of BLEST-ML, relative to the execution of K-means and Random Forest algorithms. Specifically, for each test dataset, we compared the time $t^*$, achieved by using the predicted block size, against the best, worst and average times obtained from all other possible partitionings. The different partitionings for this comparison were calculated using progressive powers of 2 from 2 to 64 for both the number of row and column blocks, which leads to 36 possible configurations. Furthermore, each test execution was repeated 10 times, taking the median value, in order to get a robust measure of execution time, preventing the evaluation process from being biased by noisy measures. Achieved results, in terms of makespan ratio and percentage makespan reduction, averaged on all test datasets, are summarized in Table 4 and discussed in the following.

| *Metric* | *Best time* | *Average time* | *Worst time* |
|---|---|---|---|
| Makespan ratio | $0.99 \pm 0.02$ | $1.25 \pm 0.06$ | $2.11 \pm 0.08$ |
| Makespan reduction (%) | $-0.79\% \pm 0.03$ | $24.71\% \pm 0.05$ | $55.06\% \pm 0.06$ |

Table 4: Average values of makespan ratio and percentage makespan reduction obtained from executing K-means and Random Forest algorithms on the synthetic test datasets.

By comparing $t^*$ with the best time measured by trying all possible partitionings, it can be noticed that the data partitioning suggested by the learning algorithm is almost always the best one, i.e. it guarantees an execution time very close to the shortest obtainable time. In particular, we measured a very little difference compared to the best execution time, with a makespan ratio almost equal to 1, and a negligible increase of execution time less than 0.8%. Regarding the comparison with the average time, we obtained a good performance improvement, with a percentage reduction of makespan equal to 24.71% and a makespan ratio equal to 1.25. These results show how the choice of an unsuitable block size may lead to a degradation of performance, which can be avoided with the aid of the proposed methodology.

We further stressed this aspect by comparing $t^*$ with the worst execution time, achieving a remarkable reduction of makespan equal to 55% and a makespan ratio equal to 2.11. Measured values confirm the ability of BLEST-ML in supporting the execution of machine learning algorithms in parallel and distributed environments.

To make more detailed and clear the benefits brought by the use of our methodology, Figure 4 and 5 show the results achieved with K-means and Random Forest in three possible cases, in which the number of rows $n$ and columns $m$ can be equal or very imbalanced. A synthetic test dataset for each case was generated as follows:

- $n >> m$: 500,000 rows, 1000 columns.
- $m >> n$: 1000 rows, 500,000 columns.
- $n \approx m$: 10,000 rows, 10,000 columns.

Again, we compare execution times achieved by using the predicted partitioning and all other possible partitionings, set using progressive powers of 2 from 2 to 64 for both the number of row and column blocks (as explained above). By observing Figure 4, we can see that, even in the presence of a high imbalance, the algorithm always suggests a block size value very close or equal to the best one, thus allowing an efficient execution of the K-means algorithm. Particularly, the time obtained by using the predicted block size is marked by the cyan circle, while the best one is marked by the green star. Moreover, the heatmap is useful to show how the variation of the block size affects the execution time, depicted in different gradations of red, where a greater intensity corresponds to a higher execution time. We observed that the time $t^*$ obtained by using the predicted block size leads to the second best time in the first two cases, and to the best time for the last one. The mean percentage difference between $t^*$ and the best time is almost equal to 1%, which shows how the partitioning suggested by BLEST-ML is a very good estimate of the optimal one. Moreover, by comparing $t^*$ with the average and worst times, we measured an average makespan ratio of 1.17 and 1.53 and an average percentage improvement of makespan equal to 14.27% and 34.44%.

The good results achieved with K-means are confirmed by the experiments performed on Random Forest, shown in Figure 5. In this case, $t^*$ resulted in the best execution time in two cases out of three (the first and the third), and the

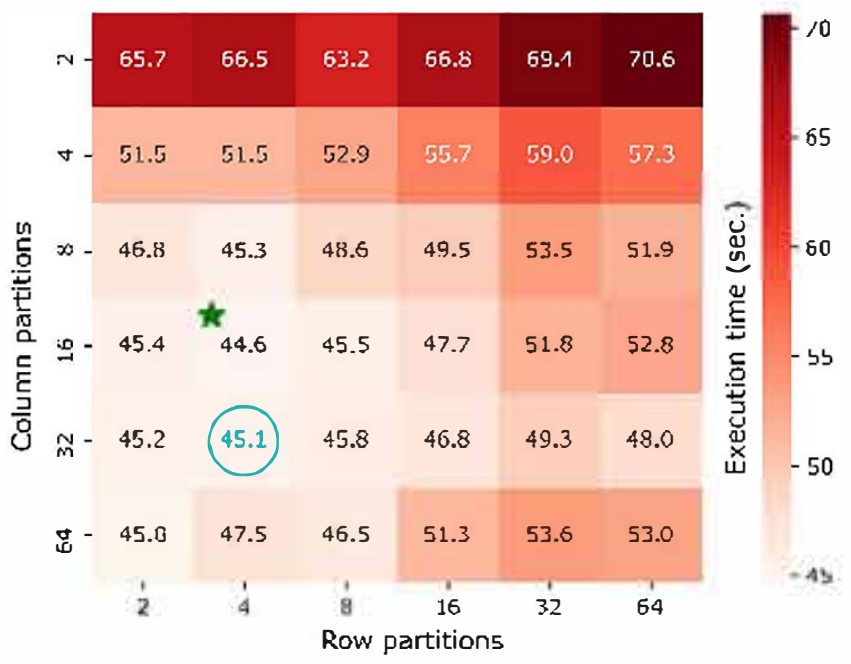

(a) K-means, $n >> m$.

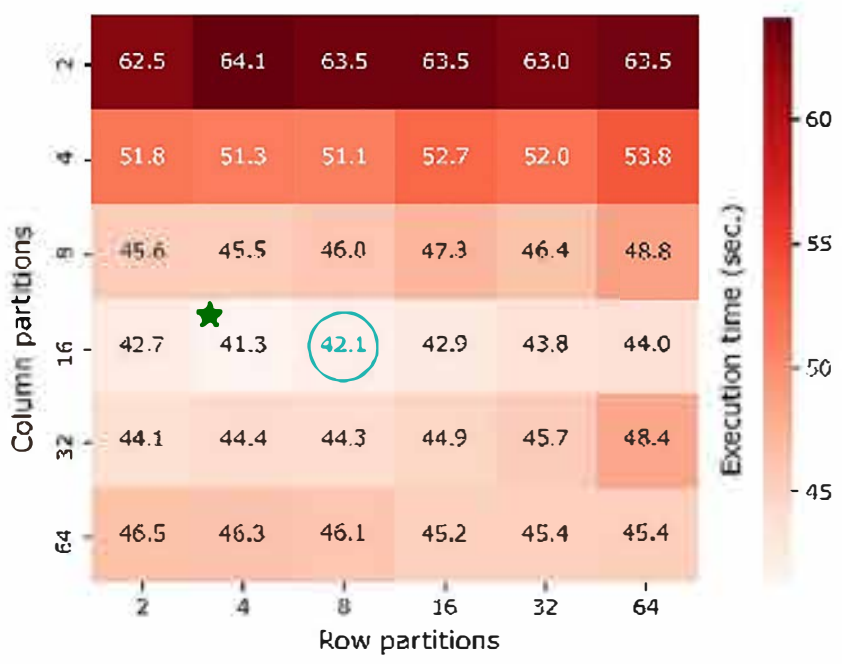

(b) K-means, $m >> n$.

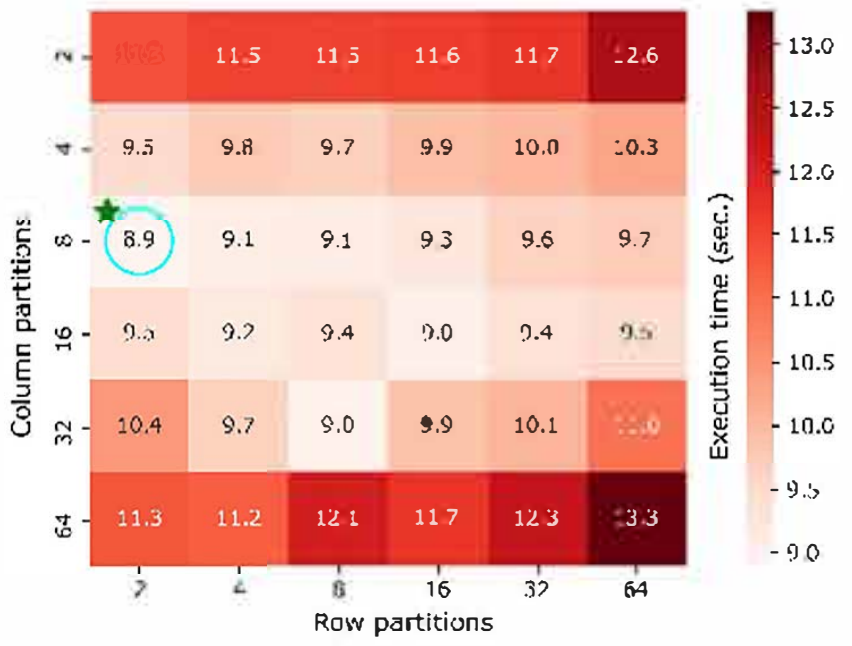

(c) K-means, $n \approx m$.

Figure 4: Execution times (in seconds) achieved by running K-means on datasets of both balanced and imbalanced shape. The time obtained with the predicted block size is marked by the cyan circle, while the best one by the green star.

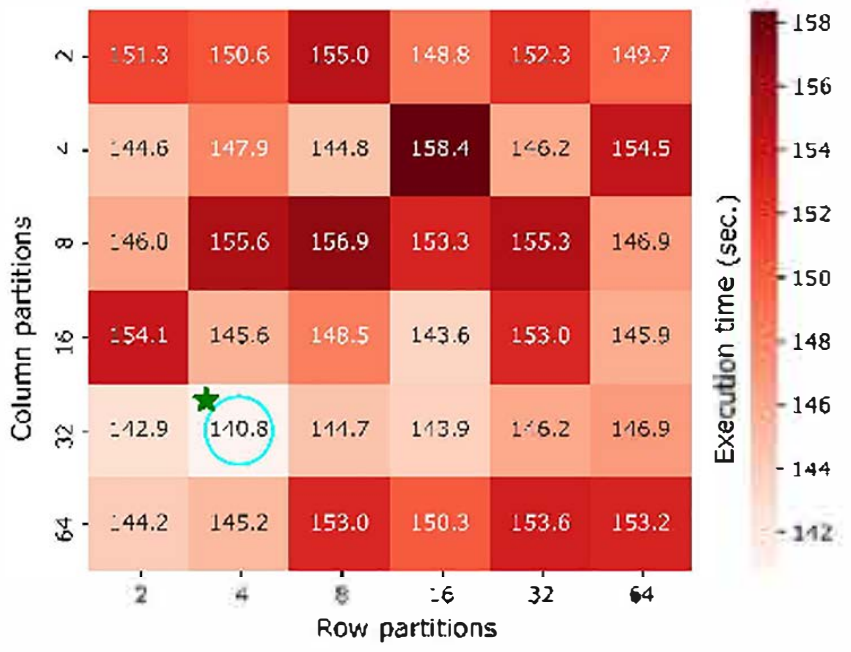

(a) Random Forest, $n >> m$.

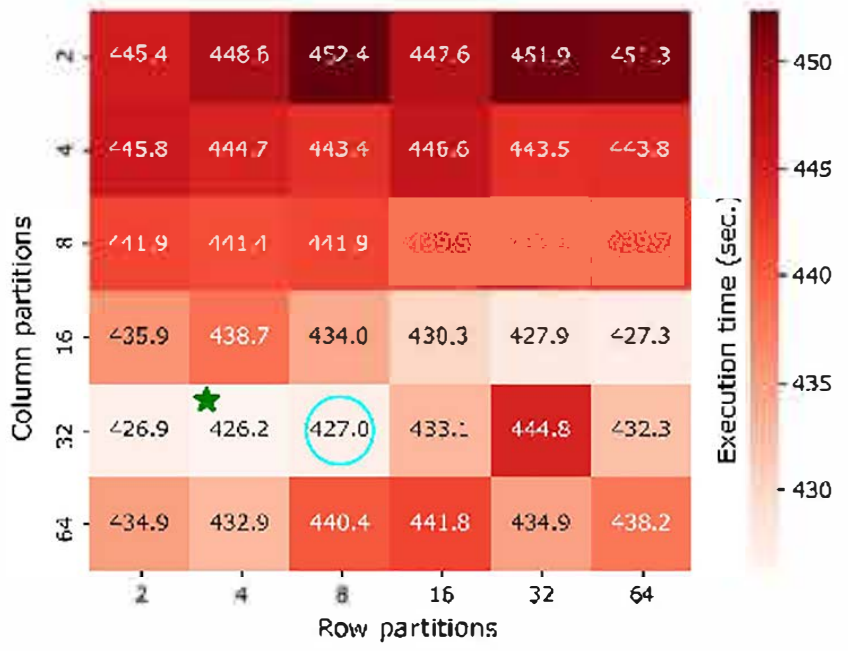

(b) Random Forest, $m >> n$.

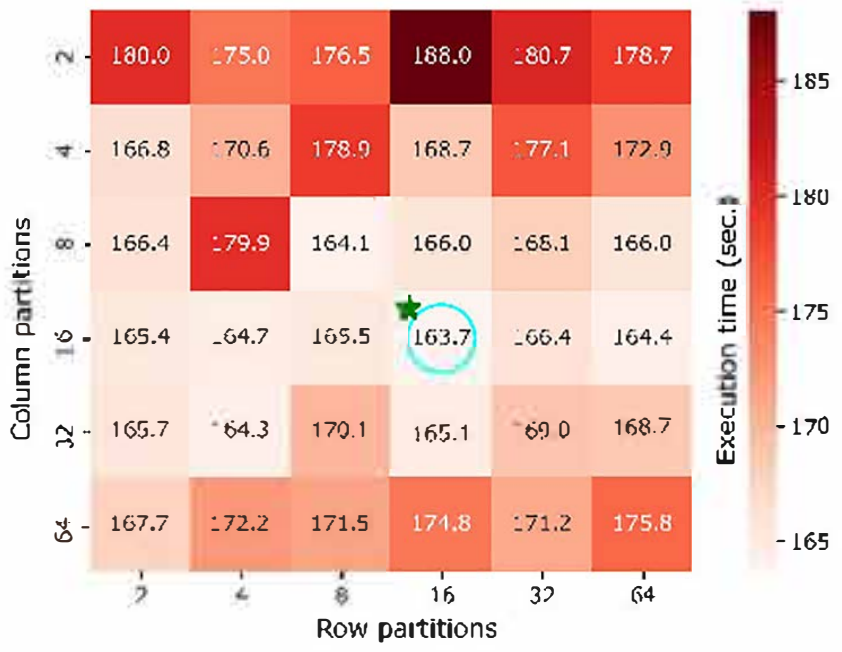

(c) Random Forest, $n \approx m$.

Figure 5: Execution times (in seconds) achieved by running Random Forest on datasets of both balanced and imbalanced shape. The time obtained with the predicted block size is marked by the cyan circle, while the best one by the green star.

third best time in the remaining case (the second). In particular, we measured a negligible difference of 0.56% between $t^*$ and the best execution time. Moreover, by comparing $t^*$ with the average and worst times we observed a makespan ratio of 1.03 and 1.10 and a percentage makespan reduction of 3.74% and 9.44%. All these results are summarized in Table 5.

| *Algorithm* | *Metric* | *Best time* | *Average time* | *Worst time* |
|---|---|---|---|---|
| K-means | Makespan ratio | $0.99 \pm 0.03$ | $1.17 \pm 0.05$ | $1.53 \pm 0.07$ |
| | Makespan reduction (%) | $-1.03\% \pm 0.02$ | $14.27\% \pm 0.04$ | $34.44\% \pm 0.06$ |
| Random Forest | Makespan ratio | $0.99 \pm 0.01$ | $1.03 \pm 0.05$ | $1.10 \pm 0.05$ |
| | Makespan reduction (%) | $-0.56\% \pm 0.04$ | $3.74\% \pm 0.06$ | $9.44\% \pm 0.09$ |

Table 5: Average makespan ratio and percentage makespan reduction measured by running K-means and Random Forest on datasets of both balanced and imbalanced shape.

To further assess the generalization abilities of the trained model, we analyzed how well it can handle datasets of increasing size without any retraining, using the same experimental setup described at the beginning of this Section. As suggested in the literature [30], we used synthetic test datasets of increasing size (i.e., 1k, 50k, 250k, 500k) to compare execution times achieved by K-means and Random Forest using the predicted block size against the best, worst, and average times. It is worth noticing that the test datasets were generated such that the resulting pairs dataset-algorithm were unknown to the model, i.e. not included in the training set. Obtained results, shown in Figure 6, highlight the ability of BLEST-ML to cope with datasets of increasing size, leading to very similar execution times compared to the best block size. Moreover, we measured marked differences in execution time compared to the average and worst times for both algorithms, with an average makespan ratio and reduction up to 2.47 and 50.4%, respectively, which further confirms the effectiveness of our methodology.

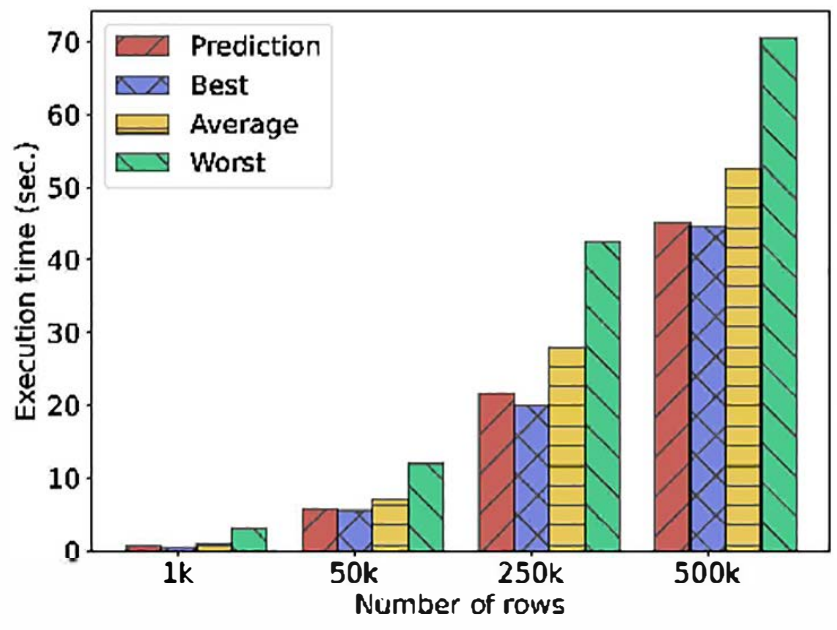

(a) K-means

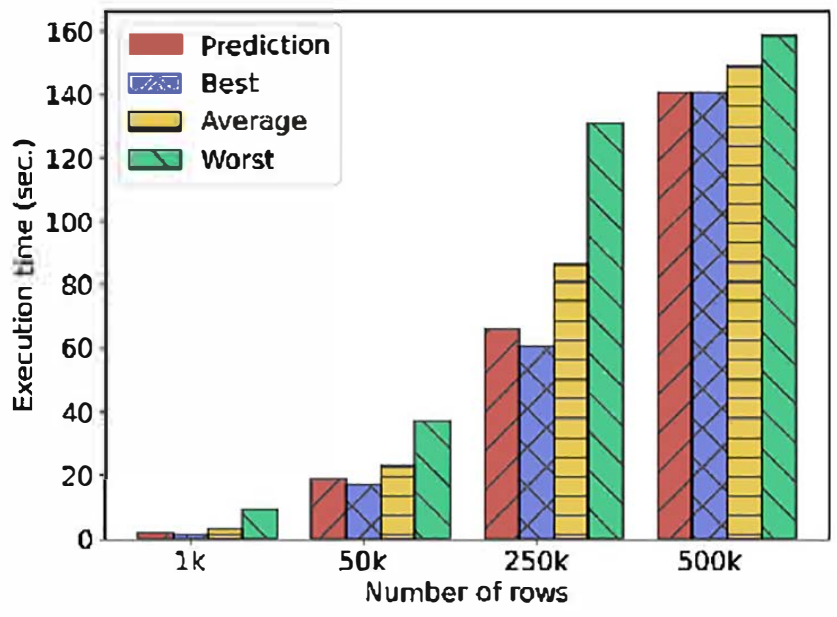

(b) Random Forest

Figure 6: Execution times (in seconds) achieved by K-means and Random Forest executed on datasets of increasing size. Time achieved by using the predicted block size is compared against the best, worst, and average times.

*5.2. Multi-node experiments*

We further investigated the effectiveness of BLEST-ML in a distributed execution environment. Specifically, we leveraged the MareNostrum 4 supercomputer (MN4) [8], located at the Barcelona Supercomputing Center. Its current peak performance is 11.15 Petaflops and it is composed of 3456 nodes, each of which has two Intel®Xeon Platinum 8160 (24 cores at 2,1 GHz each) and 96 GB of main memory. It has also 100 GB Intel®Omni-Path Full-Fat Tree Interconnection, and 14 PB of shared disk storage managed by the Global Parallel File System [31].

In this experimental evaluation, we focused on the execution of the Principal Component Analysis (PCA) algorithm. It is a dimensionality reduction algorithm, whose aim is to compute a meaningful low-dimensional representation of the input data by projecting each sample onto only the first few principal components. The log used for the extraction of the training data was enriched using several real-world datasets, listed in the following, which belong to different fields, ranging from medicine to particle physics.

- *Diabetes*: medical data, used for predicting whether or not a patient has diabetes, based on diagnostic measurements.
- *Cleveland*: medical data, used for predicting the heart disease risk based on clinical measurements.
- *Banknote*: high-resolution images, used for evaluating if a banknote is authentic or forgery.
- *Superconductors* [32]: superconductors data, used for predicting the critical temperature.
- *Accelerometer* [33]: accelerometer data, used for predicting motor failures.
- *1ubq.bck.10.crd*, *1ubq.bck.1.crd*, *1ubq.heavy.1.crd*: particle physics datasets, containing up to 1 million atom trajectories described by a varying number of features and obtained from GROMACS simulations [34].

For the experimental evaluation, we used three test datasets containing biomolecular simulation data in a *mdrcd*[2] format, containing trajectories of atoms, whose number ranges from almost 7000 to more than $30,000$. These datasets are described in Table 6.

For the execution of our experiments, we employed 16 nodes of the MN4 supercomputer, with 96 GB of RAM per node. In addition, the number of used cores per PyCOMPSs task with the large and extra-large datasets was set to 24 due to the heavy computation and their big memory size, while for the medium dataset, we used 8 cores per PyCOMPSs task.

Unlike the experiments shown in section 5.1, we did not consider the large set of all possible partitionings, as the huge size of test datasets could have led to an excessively expensive process, due to time-consuming and resource-intensive computation. In this case, instead, we compared the time achieved by using the predicted partitioning $(p_r^*, p_c^*)$ against the best partitioning that was individuated by domain experts $(\hat{p_r}, \hat{p_c})$, by following a trial-and-error approach. The obtained results are shown in Table 7.

[2]Amber trajectory format, https://ambermd.org/FileFormats.php

| *Algorithm* | *Dataset name* | *Dataset rows* | *Dataset columns* |
|---|---|---|---|
| PCA | Traj_medium | 60,000 | 20,736 |
| PCA | Traj_large | 100,000 | 59,544 |
| PCA | Traj_xlarge | 100,000 | 94,896 |

Table 6: Test datasets used to evaluate the benefits brought by BLEST-ML on the execution of the PCA algorithm on the MN4 supercomputer.

| *Dataset* | *Predicted partitioning* | | | *Manual partitioning* | | |
|---|---|---|---|---|---|---|
| | $p_r^*$ | $p_c^*$ | *Time (s)* | $\hat{p_r}$ | $\hat{p_c}$ | *Time (s)* |
| Traj_medium | 4 | 16 | 270 | 6 | 21 | 484 |
| Traj_large | 8 | 40 | 1123 | 14 | 36 | 1096 |
| Traj_xlarge | 8 | 48 | 1770 | 14 | 48 | 1825 |

Table 7: Results obtained in MareNostrum 4 using model predictions and domain expert estimates.

By comparing the time achieved by using the block size predicted by BLEST-ML with that estimated by the domain experts, we achieved quite good results, with an average value of makespan ratio and makespan percentage reduction equal to 1.27 and 14.92%, respectively. In addition, as reported in Table 7, the data partitioning suggested by BLEST-ML is the best one in two out of three test cases, resulting in the shortest execution time. Moreover, it is worth noticing that in the remaining case, corresponding to the *Traj_large* dataset, the relative difference between the two execution times is quite small ($\approx 2\%$). The prediction is indeed reasonably good, as it can be calculated quickly without involving any trial-and-error approach and requiring a small amount of resources and domain knowledge.

For the sake of completeness, in Figure 7 we provide the execution times measured by executing the PCA algorithm on the *Traj_medium* dataset with all possible partitionings, computed using progressive powers of 2, from 2 to 64, leading to 36 possible configurations. Consequently, the maximum number of blocks that can be generated is equal to $64^2$, i.e. the $64 \times 64$ configuration is selected and the dataset is divided into 64 blocks along both rows and columns. Each of the generated blocks will be then handled by PyCOMPSs tasks, one for each block.

The plot shows that the partition suggested by BLEST-ML is the third best possible. Nevertheless, while not leading to the minimum execution time, the estimate is still an excellent approximation, obtainable in a very efficient way. Indeed, our methodology allows obtaining a suitable estimate very quickly, since the machine learning model takes just a few milliseconds to compute a prediction. On the contrary, according to the execution times reported in Figure 7, a full exhaustive search would take more than 9 hours. To further stress this concept, we tuned the PCA application by using the OpenTuner [7] autotuning framework, comparing the obtained results with those achieved by BLEST-ML and the exhaustive search (see Table 8). In particular, we performed 15 tuning runs with OpenTuner, selecting the one that had the best tuning time, i.e., it identified the optimum in the shortest time. The best tuning time recorded was approximately 3 hours, finding the optimum with the UniformGreedyMutation search technique and 12 performed tests. This results in a clear improvement over exhaustive search, reducing tuning time to less than a third. Nonetheless, the time required to determine the recommended partitioning in response to a user request remains significantly greater than that employed by BLEST-ML, which uses machine learning to directly calculate an appropriate block size. Consequently, our approach can significantly improve the time-to-solution as compared to experimenting with multiple potential configurations when a user submits a request.

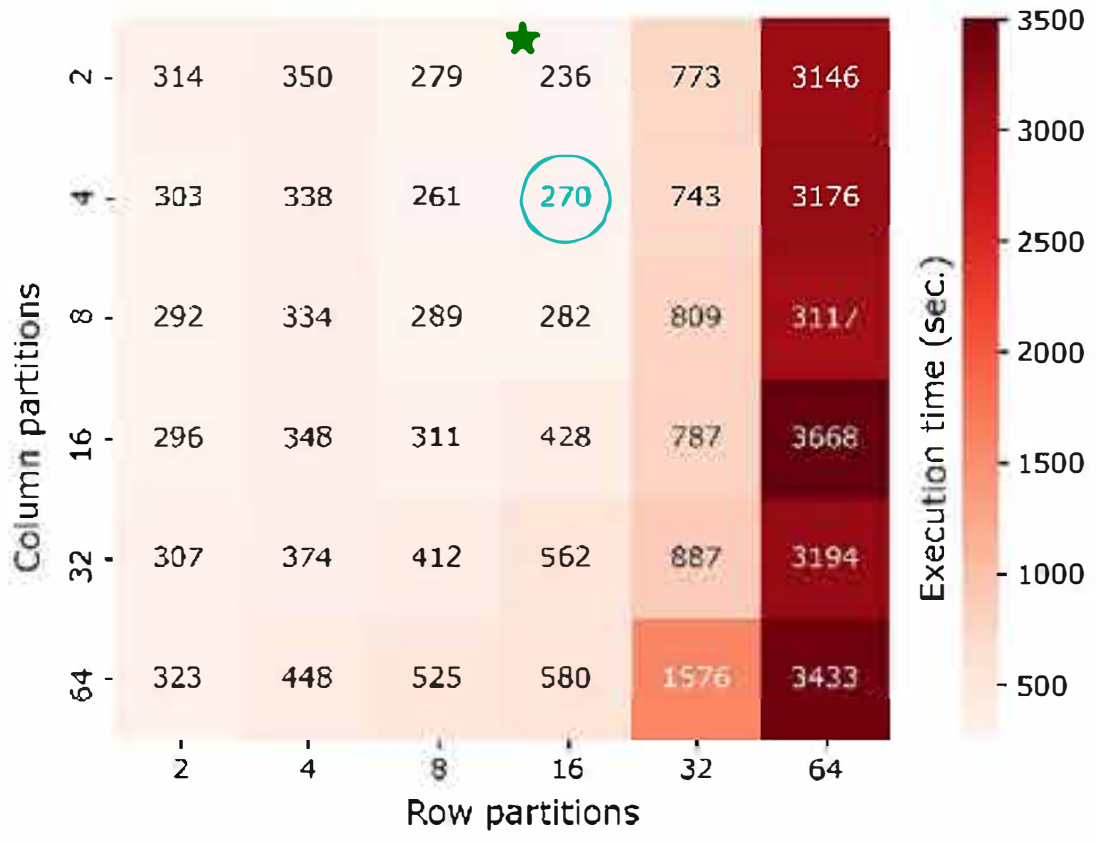

Figure 7: Execution times (in seconds) achieved in MareNostrum 4 by running PCA on the *Traj_medium* dataset. The time obtained with the predicted block size is marked by the cyan circle, while the best one by the green star.

In addition, we compared the time obtained with the predicted partitioning with the average and worst times shown in Figure 7. We measured a makespan ratio equal to 3.54 and 13.59 and a percentage reduction of makespan of 71.75% and 92.64% compared to the average and worst times. All of these results further confirm how crucial it is to choose a suitable partitioning for running data-intensive applications in high-performance distributed environments.

Finally, it is worth noticing that attempting to manually explore a set of potential configurations - or even all, in the case of the exhaustive search - in search of the one that minimizes execution time may be worthless. Indeed, once the first configuration is tested, the outcome of the executed algorithm is already available, which would make testing subsequent candidate configurations superfluous. Moreover, the optimal block size found via a manual approach would not be applicable to different unseen executions. On the contrary, the use of a machine learning model allows for direct inference and multiple re-use, as it is trained once (as described in Section 3.3) and can be used multiple times for block size prediction, given different unseen input instances.

| *Tuning technique* | *Tuning time* | *Number of tests* | *Best configuration* | *Best time (sec)* |
|---|---|---|---|---|
| Exhaustive search | 9 h 23 min | 36 | (2, 16) | 236 |
| OpenTuner [7] | 2 h 51 min | 12 | (2, 16) | 236 |
| **BLEST-ML (ours)** | $2 \cdot 10^{-3}$ sec | – | (4, 16) | 270 |

Table 8: Performance obtained by exhaustive search, OpenTuner, and BLEST-ML in tuning the PCA algorithm running on the MN4 supercomputer on the *Traj_medium* dataset.

## 6. Conclusions and final remarks

Data-intensive applications are widespread in several domains, such as bioinformatics, high-energy physics, and the modeling of natural phenomena. In such applications, an effective strategy for data partitioning is crucial to enable their efficient execution in distributed HPC environments. This paper introduced a novel methodology, namely BLEST-ML (BLock size ESTimation through Machine Learning), aimed at optimizing the execution of such applications by determining the best block size to be used for data partitioning. Our methodology was evaluated on the dislib library of PyCOMPSs, considering different execution environments, including the MareNostrum 4 supercomputer, and different real-world datasets. Experimental results show how BLEST-ML can lead to a significant improvement in application performance and a reduction in execution time, by following a machine learning-based approach. In future work, we plan to improve our methodology to make it even more generic and usable, supporting the choice of other parameters required to configure a distributed environment. Furthermore, we can investigate its applicability to frameworks and libraries other than PyCOMPSs and dislib, as it can be exploited in any case where data partitioning is essential to improve application performance and scalability.

## Ethics approval and consent to participate

Not applicable.

### Availability of data and materials

An implementation of BLEST-ML is publicly available on GitHub at the following link `https://github.com/eflows4hpc/dislib-block-size-estimation`.

### Competing interests

The authors declare that they have no competing interests.

### Funding

Not applicable.

### Authors' contributions

All the authors contributed to the structuring of this paper, providing critical feedback and helping shape the research, analysis, and manuscript. R. Cantini, F. Marozzo, and J. Ejarque conceived the presented idea and organized the manuscript. R. Cantini, F. Marozzo, A. Orsino, and F. Vazquez wrote the manuscript with input from all authors and implemented and tested the methodology. D. Talia, P. Trunfio, and R. M. Badia were involved in planning the work and supervised and reviewed the structure and contents of the paper.

### Authors' information

R. Cantini is a researcher in computer engineering at the University of Calabria. F. Marozzo is an assistant professor of computer engineering at the University of Calabria. A. Orsino is a Ph.D. student of computer engineering at the University of Calabria. D. Talia is a professor of computer engineering at the University of Calabria and an adjunct professor at Fuzhou University. P. Trunfio is an associate professor of computer engineering at the University of Calabria. R. M. Badia is the manager of the Workflows and Distributed Computing research group at the Barcelona Supercomputing Center. J. Ejarque is a senior research engineer at the Barcelona Supercomputing Center. F. Vasquez is a junior research engineer at the Barcelona Supercomputing Center.

### Acknowledgements

This work has been partially supported by the European Commission through the Horizon 2020 Research and Innovation program and the EuroHPC JU under contract 955558 (eFlows4HPC project) and by MCIN/AEI/10.13039/501100011033 and the European Union NextGenerationEU/PRTR (PCI2021-121957 and CEX2021-001148-S) and by the Spanish Government (PID2019-107255GB), and Generalitat de Catalunya (contract 2021-SGR-00412). We also acknowledge financial support from "National Centre for HPC, Big Data and Quantum Computing", CN00000013 - CUP H23C22000360005, and from "FAIR – Future Artificial Intelligence Research" project - CUP H23C22000860006.

### Consent for publication

Not applicable.